\RequirePackage{amsmath}

\documentclass[svgnames,a4paper]{llncs}
\usepackage[hyphens]{url}
\usepackage[caption=false]{subfig}
\usepackage{graphicx}
\usepackage{colortbl}
\usepackage{cprotect}
\usepackage{multirow}
\usepackage[T1]{fontenc}
\usepackage{hyperref}
\usepackage{varioref}
\usepackage{xspace}
\usepackage{paralist}
\usepackage{fancybox}
\usepackage[svgnames]{xcolor}
\usepackage{soul}
\usepackage{calc}
\usepackage{verbatim}
\usepackage{marginnote}
\usepackage{todonotes}

\usepackage[utf8]{inputenc}
\usepackage{hyperref}
\usepackage{chngcntr}
\usepackage{float}
\usepackage{enumitem}
\usepackage{setspace}
\usepackage{pgf-pie}
\usepackage{csvsimple}

\usepackage{booktabs}
\usepackage[scaled]{helvet}

\usepackage{times}
\usepackage{listingsVDM}

\definecolor{mygray}{rgb}{0.5,0.5,0.5}

\begin{document}

\title{Translating a VDM Model of a Medical Device into Kapture}
\titlerunning{}

\author{Joe Hare\inst{1} \and
  Leo Freitas\inst{1} \and
  Ken Pierce\inst{2}
}
\institute{School of Computing, Newcastle University, United Kingdom \\\email{j.hare2@ncl.ac.uk,leo.freitas@ncl.ac.uk,kenneth.pierce@ncl.ac.uk}
}

\maketitle
\begin{abstract}
As the complexity of safety-critical medical devices increases, so does the need for clear, verifiable, software requirements. This paper explores the use of Kapture, a formal modelling tool developed by D-RisQ, to translate an existing formal VDM model of a medical implant for treating focal epilepsy called CANDO. The work was undertaken without prior experience in formal methods. The paper assess Kapture's usability, the challenges of formal modelling, and the effectiveness of the translated model. The result is a model in Kapture which covers over 90\% of the original VDM model, and produces matching traces of results. While several issues were encountered during design and implementation, mainly due to the initial learning curve, this paper demonstrates that complex systems can be effectively modelled in Kapture by inexperienced users and highlights some difficulties in  translating VDM specifications to Kapture.
\end{abstract}

\section{Introduction}
\label{sec:intro}

As medical technology advances, the development of complex medical devices has become increasingly common~\cite{Neuman2012}. With complex systems comes the challenge of managing large and intricate codebases, where errors are frequently present~\cite{sommerville2011largescalecomplexsystems}. In the context of safety-critical medical devices, these errors have the potential to be particularly problematic, as they could have serious consequences for patients' health. A major cause for errors or unintended behaviour in systems is the presence of poorly defined or ambiguous system requirements~\cite{Muhammed2019}. In addition to being a safety risk, poor requirements are also a source of high development costs, as they increase the likelihood of encountering unforeseen implementation issues and lead to a greater number of defects that must be identified and resolved later in the development process.

Formal methods, such as the  Vienna Development Method (VDM), can offer a robust solution to these challenges by enabling the modelling of systems thoroughly and mathematically, allowing for the creation of system requirements whose correctness can be verified~\cite{Hu2009}, however the process of using formal methods to model complex systems can often be a similarly complex and specialist task in itself. Kapture is a tool developed by D-RisQ Ltd'\footnote{D-RisQ Ltd website: \url{https://www.drisq.com/}.} that aims to address the entry barrier to formal modelling by allowing users to link natural-language requirements to formal definitions. 

The CANDO project (Controlling Abnormal Network Dynamics using Optogenetics) explored creation of an implant for treating focal epilepsy\footnote{CANDO website: \url{http://www.cando.ac.uk/}}. This is an example of a safety-critical device, for which a VDM model was built. In this paper, we describe initial attempts to translate this VDM model to Kapture, as a valuable opportunity to explore the usability and effectiveness of Kapture. The work was undertaken primarily by the main author who had no prior experience with VDM, Kapture, nor formal methods in general, and we discuss the  challenges presented, as well as the time required to complete the model and various aspects of it.

In the remainder of this paper, Section~\ref{sec:back} provides relevant background on VDM, Kapture; Section~\ref{sec:cando} covers the CANDO project through the VDM model. Section~\ref{sec:kapture} describes the main efforts in translation to Kapture; Section~\ref{sec:discussion} provides some reflections on the process; and finally, Section~\ref{sec:conc} provides conclusions and future work.
\section{Background}
\label{sec:back}

\subsection{Safety and Verification of Complex Medical Devices}
\label{sec:back:medical}

Modern medical devices commonly rely on software systems consisting of thousands of lines of code, with software in pacemakers and defibrillators containing 80,000--100,000 lines of code~\cite{Ziang2016}. As code bases grow in size and complexity, the risk of errors and unexpected behaviour increases. Failures of safety-critical medical devices, such as the Therac-25 radiation-therapy machine that caused serious injury and death in patients~\cite{Leveson1993}, are unacceptable.

Despite new and updated standards for medical software development, such as as IEC 62304~\cite{LHF62304}, software issues in safety-critical medical devices still occur. For example, Tandem Diabetes Care's Apple iOS application, used to control an insulin pump, was rapidly draining the pump's battery, shutting it down sooner than expected, leading to 224 reported injuries ~\cite{FDATandem}.

CANDO's cortical implant is an example of a safety-critical medical device, so it is imperative that the device functions as intended. Failure of the device may mean it fails to prevent seizures in a patient. As the implant produces light, of which heat is a by-product, another possible concern is whether the device failing to turn the light off as intended could result in temperature changes affecting neuronal processes~\cite{Owen2019}. 

\subsection{VDM}
\label{sec:back:vdm}

VDM is a long-established formal method for modelling and verifying computer systems, used in frequently industry. As a model-oriented language, VDM-SL represents systems through data, state, and behaviours that act on these. Invariants on types / state allow for modelling of data, while behaviour can be defined through functions and operations, with pre- and post-conditions capturing assumptions and commitments, as well as explicit definitions that allow for execution of a subset of models.

Formal modelling with VDM can help to recognise elements of system's requirements or specification that may be ambiguous or incomplete~\cite{ISOVDM95}. Development of the existing CANDO project specification in VDM confirmed several issues with unused states and states which could not correctly self-loop in the defined Finite State Machine (FSM)~\cite{Wooding2019}.

\subsection{Kapture}
\label{sec:back:kapture}

Kapture is a tool developed by D-RisQ with the goal of lowering development costs for safety-critical software. It aims to help developers write clear, verifiable, and intuitive requirements in English that are converted into Communicating Sequential Processes (CSP). CSP-based modelling and verification has been successfully applied to medical device software previously, for example at Phillips Healthcare, where Analytical Software Design (ASD), a design approach using CSP, was used for the design and verification of a trolley system for patients to lie on~\cite{Osaiweran2010}. 

Kapture supports compliance with safety standard including DO-178C and DO-333, covering formal development for airborne systems~\cite{jacklin2012,Cofer2014}. While these standards do not officially apply to medical devices, the thorough requirements definition, verification, and traceability set out by these standards can still be helpful for strengthening the reliability and safety of a similarly critical medical device. The work reported in the paper is part of a larger project applying Kapture to medical and automotive case studies,

\section{Existing CANDO VDM Model} 
\label{sec:cando}

The CANDO project aimed to develop a cortical implant to treat epilepsy. Epilepsy is currently an incurable condition, affecting around 50 million people worldwide~\cite{WHOEpilepsy}, which causes seizures. 
As a treatment, CANDO proposed to use genetically modified neurons~\cite{Iseri2017} to make brain tissue sensitive to light, then stimulate these using an implant with \emph{optrodes} generating light induce electrical currents to influence brain activity. The implant is a closed-loop system, constantly monitoring brain activity in real-time to adjust stimulation. The brain's electrical signals are captured by electrodes and processed on a microcontroller located on the chest, connected via a wire, to control stimulation through LEDs, creating a continuous feedback loop modulating and monitoring neural activity to prevent seizures from developing~\cite{Firfilionis2021}.



\begin{figure}[tb]
    \centering
    \includegraphics[width=0.5\linewidth]{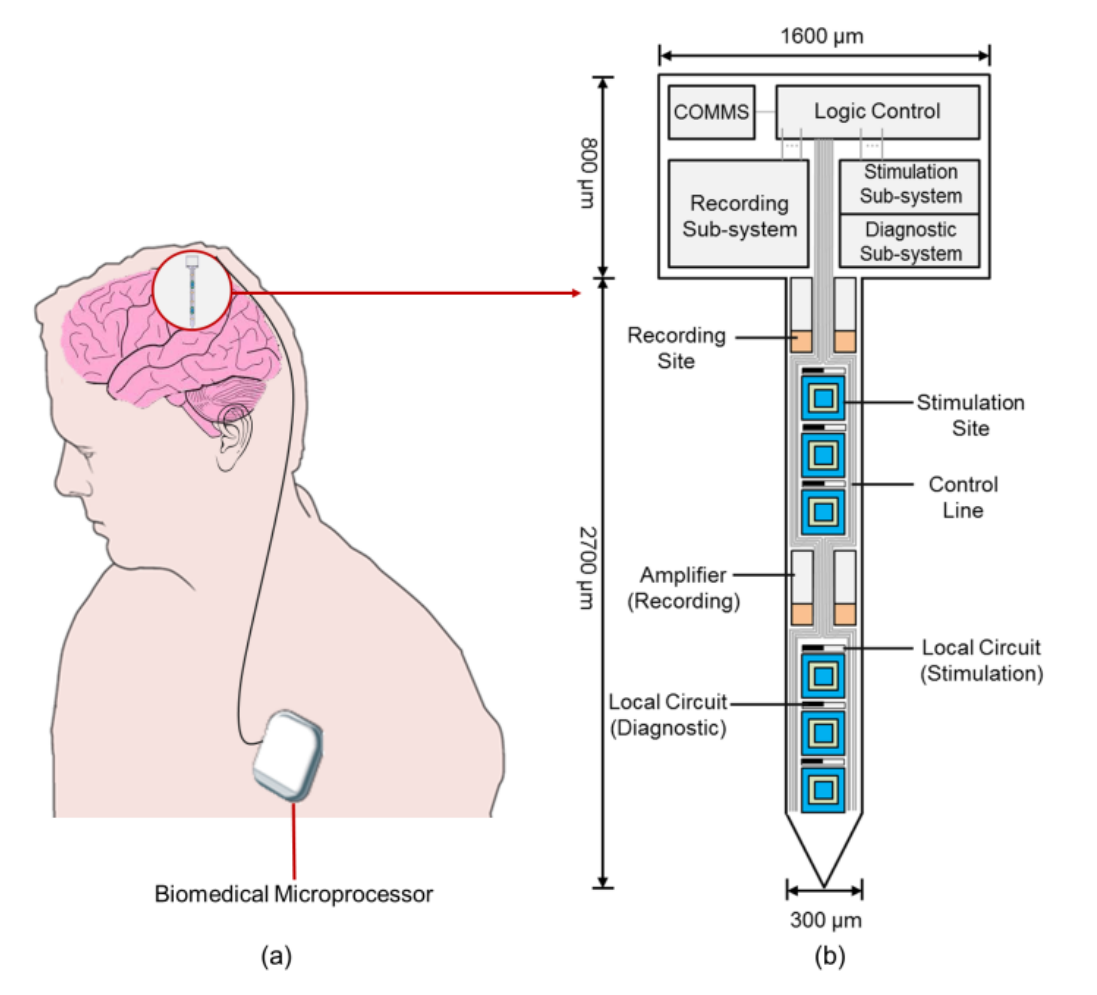}
    \caption{CANDO optrode and chest piece~\cite{Zhao2015}}
    \label{fig:optrode-diagram}
\end{figure}

The software of the CANDO system has a set of optrodes that act as both sensors and actuators of electrical activity. The behaviour of each optrode is controlled by a Finite State Machine (FSM), executed in a control loop, using a state transition table to determine state changes based on the current state and an event variable. States have their own associated function, determined by a lookup table, which perform the necessary action and updates the event variable accordingly~\cite{Pollitt2018}. 

The VDM model has approx. 1000 LOC, so it is not described in full here. The optrodes are controlled by sending and receiving packets over a bus, captured by the following types:

\begin{vdm}
types
  Packet_Data = <LED_addr> | <NO_LED_ADDR> | <DAC_value> | 
  <diag_delay> | <mem_len> | <constructed_data> | 
  <fs_ratio_to_clk> | <rec_config>;
	
  Address = <Optrode_addr>;
	
  Packet :: addr : [Address]
             cmd : [Command]
            data : [Packet_Data];
	
  Flag = bool;
	
  Bytes = nat
  inv b == b <= PACKET_LENGTH;
	
  Count = nat
  inv t == t <= MAX_COUNT; 
		
values
  PACKET_LENGTH: nat1 = 3;
  MAX_COUNT: nat1  = 2;
\end{vdm}


\noindent The key types defining the FSM are enumerations of \texttt{Command} (there are 17 commands), \texttt{Event} (21 events), and \texttt{State} (34 states) that are used to define the FSM:

\begin{vdm}
types 
  Command = <LED_ON_C>| <LED_OFF_C>| ... | <DUMMY_C>; 
	
  Event =  <CONT> | <ERROR> | ... | <GET_CMD_E>;

  State = <get_cmd> | <LED_off> | ... |  <cmd_finish>;
\end{vdm}



\noindent There are six types of state: \emph{send}, \emph{receive}, \emph{packet creator}, \emph{stage one packet creation}, \emph{stage two packet creation}, and \emph{error} states. The \emph{StateMap} type captures state transitions, with an invariant (not shown) that restricts which state transitions are possible (for example, an error state can never transition back to a non-error state). Additional \texttt{StateMap} types are defined that may only contain one of the six state types (\texttt{IdMap}, \texttt{ErrorMap}, \texttt{PacketMap}, and \texttt{ReceiveMap}). For example, a \texttt{TStateMap} (total state map) must contain \emph{all states}. The core \texttt{FSM} type is then defined through events mapped to state transitions, and a total FSM is defined through the \texttt{TFSM} type, and CANDO is a total FSM: 

\begin{vdm}
types
  StateMap = map State to State
  inv s == ...

  ...
  
  TStateMap = StateMap
  inv sm == dom sm = {s | s:State}; 
  
  ... 

  FSM = map Event to StateMap;

  TFSM = FSM 
  inv fsm == 
    dom fsm = {s | s:State}
      and
    forall e in set dom fsm & is_TStateMap(fsm(e));
	
  CandoFSM = TFSM 
  inv fsm == ...
\end{vdm}

\noindent There are two \texttt{CandoFSM} values defined, one the \texttt{original\_initial\_fsm} at the time the model was created, and the \texttt{recommended\_fsm} which contains additional invariants discovered. The functions and operations of the model then cover the various commands to construct data packets and implement methods to exercise the FSM including sending packets, receiving packets, and resetting the system. 

To broadly illustrate the working of the FSM, the following is a valid sequence. Note that blue indicates a stage one packet creation state and red a stage two package creation state, respectively:  

\begin{itemize}[noitemsep]
    \item \textcolor{blue}{set\_vLED} → send\_packet\_6 → receive\_packet\_28 → \textcolor{red}{set\_sDac} → send\_packet\_3 → receive\_packet\_27 → cmd\_finish
\end{itemize}

\noindent This is encoded within the VDM with the following transitions:

\begin{vdm}
recommended_fsm: FSM = {
  <CONT> |-> {...,
    <set_vLED>          |-> <send_packet_6>,
    <send_packet_6>     |-> <receive_packet_28>,
    <receive_packet_28> |-> <set_sDac>, 
    <set_sDac>          |-> <send_packet_3>,
    <send_packet_3>     |-> <receive_packet_27>,
    <receive_packet_27> |-> <cmd_finish>,
    ...
  }, ...
};
\end{vdm}
\section{Translation to Kapture} 
\label{sec:kapture}

\subsection{Kapture Overview}
\label{sec:kapture:overview}

In Kapture, there are three main sections the user will spend most of their time working in: Data Dictionary, Definitions, and Requirements. An additional Assumptions area allows any additional context to be informally described. The \textbf{Data Dictionary} is where the user can define data records to be used in the Definitions and Requirements sections. When creating a data record there are four templates to choose from:

\begin{description}[noitemsep]
\item[Type] allows for definition of custom data types (building on Boolean, numeric, enumerations, and arrays).
\item[Constant] are similar to VDM values but can optionally set a minimum, maximum, and tolerance value to allow for defining fixed values with possible slight deviations as constants.
\item[Signal] definitions template is for defining variable data. They have assignable data types, and optionally minimum, maximum, and initial values.
\item[Mode] allows for definition of an enumerated list of modes and a current mode to specifically capture modal behaviour in systems.
\end{description}

\noindent The \textbf{Definitions} section allows the user to create an expression and associate it with an English text description with the Definition template. These natural language definitions can then be referenced when creating Requirements or other Definitions to help make them more readable and easier to maintain. There is also a Function Definition Template, which can return a single result based on an expression and parameters.

\begin{figure}[tb]
    \centering
    \includegraphics[width=\linewidth]{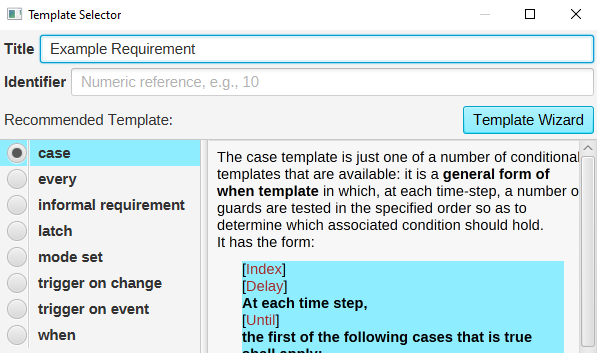}
    \caption{The available Requirement templates in Kapture}
    \label{fig:requirement-templates}
\end{figure}

The \textbf{Requirements} section allows the user to design the expected behaviour of the system by defining the conditions that should hold based on other conditions. There are several Requirement Templates to choose from as seen in the Figure~\ref{fig:requirement-templates}. The \emph{Mode Set Template} is specifically for defining the modes a component can be in at any time, as well as whether the component can be in multiple modes simultaneously or not.

The \emph{Latch} and \emph{Trigger On Change} templates are specifically for working with signals, where the latch requirement holds a signal at a certain value while a condition holds, while the Trigger On Change requires a condition holds based on a signal's value changing. For these templates more complex definitions cannot be used in place of the signals. The \emph{Trigger On Event} requires that the trigger condition is what causes the required condition to hold. The template contains extra options to specify whether the required condition must hold precisely when the trigger condition holds, or it can hold a certain amount of time after the trigger. The \emph{When} template only specifies that the required condition holds when the guard condition is true.

When writing expressions for Definitions and Requirements, there are several categories of operations to choose from to perform on the data. The most notable categories for this project are \textit{arithmetic}, \textit{comparison}, \textit{logical}, and \textit{user-defined operations}, the latter consisting of any functions created in the Definitions section. 

The \textbf{comparison} operations available for modes are \texttt{\_ is (in)active at \_}, \texttt{\_ becomes \_}, and \texttt{\_ has ever been\_}. When using the \texttt{\_ is (in)active at \_} comparison, the second field is a time-point value, which can be set to \emph{the start/end of the round}, where a \emph{round}' is an abstract tick of a wall clock. If the \texttt{\_ becomes \_} template is assumed to signal only the precise moment a mode changes to the specified status (e.g. if mode X of component A is already active, then the expression \texttt{mode "A.X" becomes active} would be \texttt{false}), then it seems that without creating extra data and definitions or using time-points, there is no way to create a guard condition to check the current status of a mode.

\begin{figure}[tb]
    \centering
    \includegraphics[width=\linewidth]{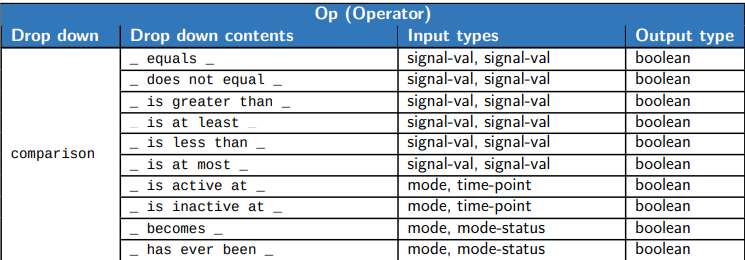}
    \caption{The available comparison operations in Kapture and their input types}
    \label{fig:kapture-comparisons}
\end{figure}

\subsection{Kapture Model}
\label{sec:kapture:model}


Given the work in this paper was carried out by a new user of both VDM and Kapture, an iterative approach was taken, breaking the overall task down into manageable pieces, with most going through several versions that were evaluated in an ongoing basis. 

\subsubsection{States and Events}

The first decision was how to represent the \emph{states} and \emph{events} of the FSM. In Kapture, it seems logical that states would be represented with modes, however as VDM-SL does not have in-built abstractions for modal behaviour, it could have been the case that defining them in this way in Kapture may have lead to problems later. Despite this, after experimenting with representing the data in multiple combinations of modes and signals, modes were used to represent the states, with events represented as enumerated values type, and a separate signal to represent the current event. 

The second decision was to decide how a state transition requirement should be defined. The logic of the requirement needed to be as such:

\begin{verbatim}
if currentState = A and event = X 
then currentState := B
\end{verbatim}

As mentioned in previous section, there is no built-in comparison operator in Kapture to check if a mode is currently active, and as such, implementing this logic would require extra steps. Two options considered were:
\begin{itemize}
    \item Implement transitions using the \texttt{\_ is active at \_} comparison. This would mean introducing the concept of rounds into the model, and would require two definitions for each state: one for when the state is active at the start of a round, and one for when it is active at the end.
    \item Create a separate signal to represent the current active state, which can be used directly in a requirement to determine the next state, then use the \texttt{\_ becomes \_} comparison to set the corresponding `mode' versions of the current state to inactive and the next state to active, then update the current state signal.  
\end{itemize}

Ultimately the former method was chosen, as after attempting to create the model using the latter method, it transpired that maintaining a separate signal to indicate the current state would introduce redundancy to the model, given that the mode component already encapsulates this information.

To do this, a \emph{states} group was created in the Definitions section with two subgroups: \emph{to states} and \emph{from states} indicating when a state is active at the start of the round and end of the round, respectively. In the Requirements section, a Requirement was created for each transition in the FSM, grouped by the required event. The \textit{Trigger On Event} requirement template was used for state transitions, as the guard for the current event and state are triggers for the next state. To address the concern of setting states to inactive after transitioning from them, an addition \textit{Mode Set} requirement was added to require the state component must have exactly one state active at a time. 

\subsubsection{FSM Translation}

The FSM in VDM uses six state types using the \texttt{StateMap} type. When performing the translation, Kapture does not allow the type of element in an Array to be a Mode, so it was not possible to create separate sets for each group in the Data Dictionary. Instead, the definitions in the \emph{to states} and \emph{from states} groups mentioned above were used, combining them in expressions using logical OR comparisons to create new definitions that hold true when any state within a state group is active at the start or end of the round. 

The invariants in the VDM model, elided in the above for brevity, establish the following constraints:

\begin{enumerate}[nosep]
    \item Constraints defined within \texttt{StateMap} invariant:
    \begin{enumerate}[label*=\arabic*., nosep, left=1em]
        \item No state can map to \texttt{start}
        \item \texttt{start} only maps to \texttt{get\_cmd} or \texttt{error\_}
        \item \texttt{chip\_rst} only maps to \texttt{get\_cmd} or \texttt{error\_}
        \item \texttt{error\_} only maps to \texttt{get\_cmd}, \texttt{error\_}, or \texttt{chip\_rst}
        \item \texttt{cmd\_finish} only maps to \texttt{error\_}
        \item Packet creator states only map to send states or \texttt{error\_}
        \item Receive states only map to stage two packet creator states, \texttt{cmd\_finish}, or \texttt{error\_}
        \item \texttt{get\_cmd} only maps to stage one packet creator states or \texttt{error\_}
    \end{enumerate}
    \item For \texttt{CONT}, all send states map to receive states
    \item For \texttt{SPI\_TX\_FINISH} send states map to themselves
    \item For \texttt{SPI\_RX\_FINISH} receive states map to themselves
    \item For event \texttt{CONT}, all packet creator states map to send states
    \item For event \texttt{CONT}, all error states map to \texttt{error\_}
    \item For event \texttt{CONT}, \texttt{start} maps to \texttt{get\_cmd}.
    \item For event \texttt{CONT}, \texttt{error\_} maps to \texttt{chip\_rst}
    \item For event \texttt{GET\_CMD\_E}, \texttt{error\_} maps to \texttt{get\_cmd}
    \item For event \texttt{GET\_CMD\_E}, \texttt{chip\_rst} maps to \texttt{get\_cmd}
    \item For event \texttt{CONT}, receive states map to stage two packet creators / \texttt{cmd\_finish}
    \item Under events \texttt{CONT}/\texttt{GET\_CMD\_E}, \texttt{error\_} remains in \texttt{error\_} unless overridden
\end{enumerate}


For 1.1, Kapture's unconditional \textit{Every} requirement template was used as the constraint unconditionally blocks states from transitioning to \texttt{\_start}. For constraints 1.2 through to 2, as well as 5, 6, and 11, the \textit{When} template was used to indicate that these requirements represent rules that must hold under certain conditions, rather than triggers for the state transition. The \textit{Trigger On Event} template was used to represent the explicitly defined state transitions in constraints 7 through to 10. 

In the VDM model, constraints 3 and 4 use \texttt{IDMap} (identity), which maps a state to itself. In Kapture, as the \textit{from state} and \textit{to state} definitions are defined separately, with no explicit link, it was not possible to easily write an expression that simply states that the state to transition from is the same as the state being transitioned to. The possibility of including signals to represent the \textit{to state} and \textit{from state} were considered, but was decided against to avoid making the Mode component redundant. Instead, new `IDMap' definitions were created for send and receive states, as these are the only states checked against IDMap in VDM, to say any \textit{from state} definition that holds implies its corresponding \textit{to state} definition also holds. 

\subsubsection{Model State}

The state of the VDM model is defined as (see type definitions above):

\begin{vdm}
state FSM of
  fsm                 : CandoFSM
  currentSt           : State
  currentEvt          : Event
  currentCmd          : Command
  command_finish_flag : Flag
  optrode_TX_finish   : Flag
  optrode_RX_finish   : Flag
  s_packet            : [Packet]
  bytes_received      : Bytes
  bytes_sent          : Bytes
  tx_cnt              : Count
inv mk_FSM(-,-,-,-,-,-,-,-,-,-,-) == ...
init FSM == ...
\end{vdm}


\noindent An enumerated values \textit{Command} type was created for the \texttt{currentCmd} type. Although \texttt{Flag} is simply a Boolean type, to keep the Kapture model close to its VDM counterpart, a Flag type was created and signals assigned to it. The \texttt{Bytes} and \texttt{Count} type are defined natural numbers with invariants defining upper bounds. In Kapture, constant records for \texttt{PACKET\_LENGTH} and \texttt{MAX\_COUNT} were created and used them for the maximum values of the Bytes and Count types respectively. These had to be defined as integers in Kapture as there is no option for natural numbers (due to this, a minimum value of 0 was also set).

\subsubsection{Functionality} The VDM model defines operations \texttt{execute} and \texttt{manual} to perform the control loop and exercise the FSM. Since Kapture focuses on requirements, it does not include the same notion of operation as VDM. However, since Kapture includes a notion of rounds, the explicit implementation of the execute operation is not required. Therefore the requirements for each transition can be encoded using the \textit{When} requirement template. Several of the VDM operations use nested conditional statements, which aren't directly supported in Kapture. To get around this issue, the requirements were remade as a single case template, splitting each possible trace of the operation in VDM into different cases. Kapture provides an HTML export to allow for models to be read outside of the GUI. To illustrate this, the first transition of the trace show previously appears as:

\begin{center}
\sethlcolor{Cyan}
\noindent\begin{tabular}{ll}
\multicolumn{2}{l}{\textcolor{Purple}{\hl{\textbf{projectId/28.00: set\_vLED to send\_packet\_6}}}} \\
If & \\
& \textcolor{blue}{\underline{The fsm is in state set\_vLED at the start of the round}} \\
\multicolumn{1}{r}{\textcolor{blue}{and}} & \textcolor{blue}{\underline{The current event is CONT}} \\
\multicolumn{2}{l}{occurs, then} \\
             & \textcolor{blue}{\underline{The fsm is in state send\_packet\_6 at the end of the round}} \\
holds.       & \\
\end{tabular}
\end{center}


\noindent Another issue that arose from operations that need to increment \texttt{bytes\_received} and \texttt{bytes\_sent}. In VDM, this increment is written as:

\begin{verbatim}
    bytes_received := bytes_received + 1;
\end{verbatim}

\noindent It was possible to encode this behaviour by defining signals like `next\_bytes\_received' and `next\_bytes\_sent' with Trigger On Event template requirements, using the optional \textit{AtSomePoint} and \textit{Within} fields, to make the fields the same within 0 rounds. This was also used for \texttt{tx\_cnt signal} and \texttt{optrode\_TX/RX\_finish flags}. This is certainly a `hack' and a more abstract version of a requirements model may well be able to handle these in a more idiomatic way in Kapture.  

\subsubsection{Pre- and Post-conditions} In the VDM model, operations are defined with both explicit functionality and associated pre- and post-conditions, allowing a combination of declarative and operational specification. Each VDM operation was captured as a requirement with the postconditions extended to consider all possible cases, stating exactly which event the system should be in and what value \texttt{tx\_cnt} should be relative to its initial value for each. This contrasts the VDM models where each  postcondition simply states the possible current events after execution. This approach effectively encodes the intended behaviour of the operation within Kapture's declarative framework.

\subsection{Testing}
\label{sec:kapture:testing}


To check the Kapture model, the VDM-SL was used to generate traces for every initial command value, with each compared to traces generated by the Kapture model with the same command. 

While the VDM traces contain just a single field for the packet, it is split into three separate fields in the Kapture traces, as this is how s\_packet is represented in Kapture. The Kapture traces also additionally include the optrode\_TX/RX\_finish flag values, as this made the trace much easier to complete. Every time a value changes in the Kapture traces, a reference to the ID of the Kapture requirement responsible for the change is listed next to it. Some values such as bytes sent and received list two different requirements next to them. These show the requirement that updates the next version of that record, and then the requirement where the actual record is updated.

The Kapture traces were completed manually by checking through all of the requirements that held for the current values of all records. Kapture's filter tool made this task significantly quicker, allowing me to filter requirements that reference the current relevant signals and states. 


\begin{figure}[tb]
    \centering
        \includegraphics[width=\textwidth]{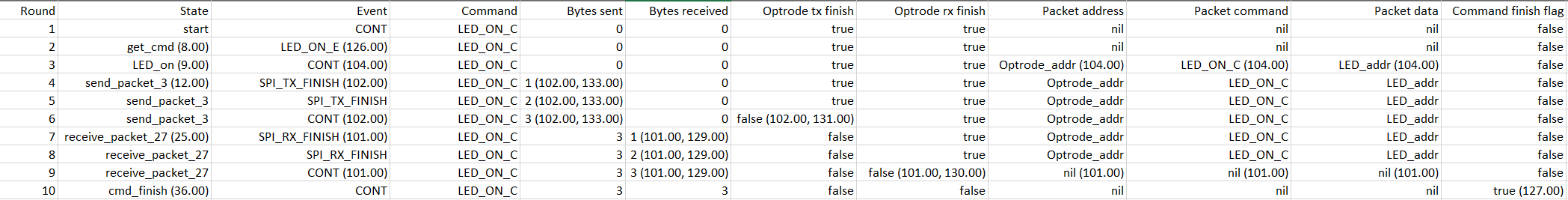}%
    \caption{A trace of the Kapture model with initial command \texttt{LED\_ON\_C}}
    \label{fig:kapture-led-on-trace}
\end{figure}

The Kapture traces are formatted slightly differently to the VDM traces. While the VDM traces contain just a single field for the packet, it is split into three separate fields in the Kapture traces, as this is how s\_packet is represented in Kapture. The Kapture traces also additionally include the optrode\_TX/RX\_finish flag values, as this made the trace much easier to complete. Every time a value changes in the Kapture traces, a reference to the ID of the Kapture requirement responsible for the change is listed next to it. Some values such as bytes sent and received list two different requirements next to them. These show the requirement that updates the next version of that record, and then the requirement where the actual record is updated.

The Kapture traces were completed manually by checking through all of the requirements that held for the current values of all records. Kapture's filter tool made this task significantly quicker, allowing me to filter requirements that reference the current relevant signals and states. 
\section{Discussion}
\label{sec:discussion}





\subsection{Model Coverage and Effort}
\label{sec:discussion:cov}

The Kapture model was completed over a time period of eight weeks, although a large majority of the implementation was completed in a shorter four-week time period. The proportion of time spent on translating into Kapture was as follows: 40\% on state operation requirements, 32\% on state transition requirements, 21\% on looping behaviour, and 7\% on FSM looping behaviour.

The propertion of the VDM model covered by the Kapture model is a difficult metric to measure precisely, as due to the differences between the two modelling methods, various parts of the VDM model could not be translated in a one-to-one direct manner. While they are functionally implemented in Kapture, drawing a link between the two equivalent parts in each model is not always possible.

Despite this, the Kapture model covers close to all parts of the VDM model. Excluding some auxiliary definitions (such as printing to console), the only parts of the VDM model not fully covered are the functions and maps that ensure all events and states are included in the finite state machine's domain, and the grouping of states by whether a packet should exist after the state's corresponding state operation is executed. When accounting for the lack of these implementations, the proportion of the VDM model covered by the Kapture model is over 90\%. In Kapture, this translates to 26 Data Dictionary records, 105 Definitions, and 113 Requirements.


\subsection{Issues Encountered During Development}
\label{sec:discussion:issues}

Using Kapture was generally an intuitive experience, but some issues were encountered with the tool during the project. As the tool uses a GUI, creation of a large number of definitions and requirements for similar cases is time consuming. There were 84 such cases in the CANDO model, where the fastest option was to create a requirement and copy repeatedly using external keyboard input automation scripts. 

During testing, it was discovered that many of the requirements in the model were created as the wrong template, with several \textit{Trigger On Event} templates that should have been \textit{When} templates, and \textit{When} templates that should have been \textit{Trigger On Event} templates. It is not currently possible to change the template of an existing requirement, so each had to be recreated manually.


Another issue was a lack of an easy method to copy records from one Kapture file to another. Mid-way through the project, a group of definitions was accidentally deleted that contained a large conditional definition. A backup was created, but the only way to copy it into the current project is by importing the old project. Kapture allows the user to select whether to import Definitions, Requirements, Data, and Assumptions separately, but there is no further control to specify exactly which of the definitions to import, so the only option was to import a large number of definitions and delete them individually to get the function back. 

\subsection{Impact of Novice User}
\label{sec:discussion:user}

Due to the lack of experience of the project team, there were several issues encountered that may only come from lack of experience with Kapture. One of the biggest challenges was deciding between Kapture's different methods of implementation for different parts of the VDM model, such as whether states, events, and commands should be modes, signals, or a combination of both; whether to use \textit{Trigger On Event} templates or \textit{When} templates for requirements where either would seem appropriate; or when to split a requirement into several smaller requirements rather than using a larger \textit{Case} template requirement or function. 

While the answers to these questions did become apparent, it often took progressing far enough with the model to run into an issue with the current implementation for it to become clear as to why one implementation was the better option, requiring time-consuming reworking and modifying earlier parts of the model.

Another difficulty came from having the system states set up as modes, while keeping all of other state variables as signals. The initial decision to implement only states as modes and all other variables as signals was mainly based on modes being likened to states in Kapture's documentation. This made updating and comparing variables confusing early on in the project, and it later became apparent that the model may have been better organised by defining state variables consistently as all modes or signals.

\subsection{Impact of VDM on the Kapture Model}
\label{sec:discussion:vdm}

Towards the end of the project, the model was sent to D-RisQ for feedback. The primary feedback was that the model is an embedding of the VDM specification into Kapture, as opposed to a higher-level set of requirements for which  Kapture is intended. The VDM specification was designed to be executable and to permit proof through Isabelle.  So while having a clear, intuitive model in VDM helped an unfamiliar user, the lower-level of abstraction seems to be a mismatch with Kapture.


Several of the issues described above stemmed from having to work with several similar requirements for each enumerated value of a data type. This may be due to the fact that Kapture is intended for the creation of high-level requirements, whereas the CANDO model is of the optrode command interface for a medical implant, with data records representing logical states and data packets. The lower-level nature of the model may have resulted in an atypical use-case of Kapture. This suggests perhaps a further iteration with guidance from D-RisQ on abstraction could be the best next step.  It would be interesting to see whether a more abstract version of the VDM model could be created and translated by another user new to Kapture.  
\section{Conclusions and Future Work}
\label{sec:conc}

In this paper we described the translation of a VDM model of a medical device, described as an FSM, to Kapture. The resulting Kapture model covers approximately 90\% of the original VDM model. The fact that the Kapture model was created by a user with no prior experience of either VDM or Kapture suggests an accessibility of both methods; completing the Kapture model over a few weeks supports D-RisQ's claims regarding Kapture's usability, while suggesting that, with minimal training, new users could become productive with Kapture in a short period of time, offering potential for cost saving in development.

The readability of the Kapture model, through the use of its English language constructs, along with the organisation provided by Kapture's different sections and the groups defined within them, makes for a model of the CANDO project which someone with little to no prior understanding of the CANDO project or VDM may be able to read and understand in less time than it's VDM counterpart, helping to make future work on the CANDO project by others more accessible.

Most problems encountered throughout development stemmed from an initial lack of understanding of formal methods and the conceptual difference between Kapture and VDM, leading to suboptimal design decisions early on in the project in an attempt to keep the Kapture model as close to its VDM counterpart as possible, which then evolved into larger issues in the later stages. Undertaking this or a similar project a second time would result in a much more efficient and streamlined development process.

There are several other opportunities for further work on this project. First, the Kapture model has an issue with the incorrect templates being used for several requirements, to better follow the Kapture idiom (as suggested by feedback from D-RisQ). While several traces have been completed to test that the model behaves correctly, the next step in testing the model would be to use D-RisQ's Modelworks tool to automatically verify requirements created in Kapture~\cite{DRisqwhitepaper}, allowing the model to make stronger claims of reliability and correctness.

During development, there was some uncertainty regarding whether modes or signals would be the best approach for representing the system's state variables, and as a result of this, the final model incorporates a combination of the two. To investigate this design choice further, it would be valuable to create two additional versions of the model: one that uses only modes for state variables, and one that relies solely on events. The requirements for each of these models could then be evaluated in comparison to each other to determine whether one approach offers a clearer advantage in terms of clarity or alignment with intended system behaviour.


%
%
%

\section*{Acknowledgements}

We acknowledge funding from the Federal Ministry for Economic Affairs and Climate Action (BMWK) on the basis of a decision by the German Bundestag through the Central Innovation Programme for SMEs (ZIM) and Innovate UK under the Third Call for Proposals for Joint Research and Development (R\&D) Projects between Germany and the United Kingdom. We would like to thank our colleagues at D-Risq for their support and feedback, including Dr Nick Taylor, Dr Colin O'Halloran, and Dr Anthony Smith. 

\clearpage
\bibliographystyle{splncs03}
\bibliography{intocps,intocpsdev,dan,references}

\newcommand{\noop}[1]{}
\begin{thebibliography}{10}
\providecommand{\url}[1]{\texttt{#1}}
\providecommand{\urlprefix}{URL }

\bibitem{ISOVDM95}
Andrews, D., Bruun, H., Hansen, B., Larsen, P., Plat, N., et~al.: {Information Technology --- Programming Languages, their environments and system software interfaces --- Vienna Development Method-Specification Language Part~1: Base language}. ISO (1995), draft International Standard: 13817--1

\bibitem{Cofer2014}
Cofer, D., Miller, S.P.: Formal methods case studies for do-333. Tech. rep., Rockwell Collins Inc. (2014)

\bibitem{Firfilionis2021}
Firfilionis, D., Hutchings, F., Tamadoni, R., Walsh, D., Turnbull, M., Escobedo-Cousin, E., Bailey, R.G., Gausden, J., Patel, A., Haci, D., Liu, Y., LeBeau, F.E.N., Trevelyan, A., Constandinou, T.G., O’Neill, A., Kaiser, M., Degenaar, P., Jackson, A.: A closed-loop optogenic platform. Frontiers in Neuroscience  15 (2021)

\bibitem{FDATandem}
Food, U., Administration, D., et~al.: Tandem diabetes care, inc. recalls version 2.7 of the apple ios t: connect mobile app used in conjunction with t: slim x2 insulin pump with control-iq technology prompted by a software problem leading to pump battery depletion. FDA  (2024)

\bibitem{Hu2009}
Hu, A.J.: Automatic formal verification of software: Fundamental concepts. In: 2009 International Conference on Communications, Circuits and Systems. pp. 1155--1159 (2009)

\bibitem{Iseri2017}
Iseri, E., D., K.: Implantable optoelectronic probes for in vivo optogenetics. Journal of neural engineering  14(3) (2017)

\bibitem{jacklin2012}
Jacklin, S.A.: Certification of safety-critical software under do-178c and do-278a. In: Infotech@aerospace Conference (2012)

\bibitem{Leveson1993}
Leveson, N., Turner, C.: An investigation of the therac-25 accidents. Computer  26(7),  18--41 (1993)

\bibitem{LHF62304}
{LHF Regulatory}: Iec 62304 update: Powerful changes you need to know and why they matter (2025), \url{https://lfhregulatory.co.uk/iec-62304-update-2026/}

\bibitem{Neuman2012}
Neuman, M.R., Baura, G.D., Meldrum, S., Soykan, O., Valentinuzzi, M.E., Leder, R.S., Micera, S., Zhang, Y.T.: Advances in medical devices and medical electronics. Proceedings of the IEEE  100 (2012)

\bibitem{Osaiweran2010}
Osaiweran, A., Boosten, M., Mousavi, M.R.: Analytical Software Design: Introduction and Industrial Experience Report. Technische Universiteit Eindhoven (2010)

\bibitem{Owen2019}
Owen, S., Liu, M., Kreitzer, A.: Thermal constraints on in vivo optogenetic manipulations. Nature Neuroscience  22 (2019)

\bibitem{Pollitt2018}
Pollitt, A.: Verifying the CANDO Project Optrode Command Interface in eCv. Master's thesis, School of Computing Science, Newcastle University, UK (2018)

\bibitem{Muhammed2019}
Riaz, M.Q., Butt, W.H., Rehman, S.: Automatic detection of ambiguous software requirements: An insight. In: 2019 5th International Conference on Information Management (ICIM). pp. 1--6 (2019)

\bibitem{sommerville2011largescalecomplexsystems}
Sommerville, I., Cliff, D., Calinescu, R., Keen, J., Kelly, T., Kwiatkowska, M., McDermid, J., Paige, R.: Large-scale complex it systems (2011), \url{https://arxiv.org/abs/1109.3444}

\bibitem{DRisqwhitepaper}
Tudor, N.: Meeting {DO-178C} and the formal methods supplement {DO-333}. White paper, {D-RisQ Ltd.} (may 2019)

\bibitem{Wooding2019}
Wooding, B.: Using Formal Methods and Proof to Verify a CANDO Epilepsy Medical Device. Master's thesis, School of Computing Science, Newcastle University, UK (2019)

\bibitem{WHOEpilepsy}
{World Health Organisation}: Epilepsy: a public health imperative. World Health Organization (2019)

\bibitem{Zhao2015}
Zhao, H., Dehkhoda, F., Ramezani, R., Sokolov, D., Degenaar, P., Liu, Y.: A cmos-based neural implantable optrode for optogenetic stimulation and electrical recording. 2015 IEEE Biomedical Circuits and Systems Conference (BioCAS)  (2015)

\bibitem{Ziang2016}
Ziang, Z., Abbas, H., Jang, K., Mangharam, R.: The challenges of high-confidence medical device software. Computer  49(1),  34--42 (2016)

\end{thebibliography}

\end{document}